# Metal/BaTiO$_3$/β-Ga$_2$O$_3$ Dielectric Heterojunction Diode with 5.7 MV/cm Breakdown Field

Zhanbo Xia[1], Hareesh Chandrasekar[1], Wyatt Moore[1], Caiyu Wang[1], Aidan Lee[3], Joe McGlone[1], Nidhin Kurian Kalarickal[1], Aaron Arehart[1], Steven Ringel[1,2], Fengyuan Yang[3], Siddharth Rajan[1,2]

[1] *Electrical & Computer Engineering, The Ohio State University, Columbus, OH 43210, USA*

[2] *Materials Science and Engineering, The Ohio State University, Columbus, OH 43210, USA*

[3] *Department of Physics, The Ohio State University, Columbus, OH 43210, USA*

**Abstract:** Wide and ultra-wide band gap semiconductors can provide excellent performance due to their high energy band gap, which leads to breakdown electric fields that are more than an order of magnitude higher than conventional silicon electronics. In materials where p-type doping is not available, achieving this high breakdown field in a vertical diode or transistor is very challenging. We propose and demonstrate the use of dielectric heterojunctions that use extreme permittivity materials to achieve high breakdown field in a unipolar device. We demonstrate the integration of a high permittivity material BaTiO$_3$ with n-type β-Ga$_2$O$_3$ to enable 5.7 MV/cm average electric field and 7 MV/cm peak electric field at the device edge, while maintaining forward conduction with relatively low on-resistance and voltage loss. The proposed dielectric heterojunction could enable new design strategies to achieve theoretical device performance limits in wide and ultra-wide band gap semiconductors where bipolar doping is challenging.

## Introduction

The potential of different materials for vertical power switching is often assessed by calculating the Baliga Figure of Merit ($\frac{V_{BR}^2}{R_{ON}} = \frac{\mu \epsilon E_c^3}{8}$) [1]. In the case of wide and ultra-wide band gap materials, the high breakdown fields and the relatively good transport properties make the BFOM significantly higher than conventional Si electronics. However, the breakdown field predicted for a material requires that the entire band gap be presented across the rectifying junction, such as in a PN junction. This is challenging to achieve in several wide band gap materials where bipolar doping is not available, or presents technological challenges. Schottky junctions can provide excellent rectification but the reverse breakdown of Schottky rectifiers is limited by the Schottky barrier height, which is significantly lower than the band gap in most wide band gap semiconductors.[2] In this work, we show, using the case of β-$Ga_2O_3$, a dielectric heterojunction approach that can enable field management so that high breakdown fields can be achieved even in unipolar junctions.

The large breakdown electric field (6-8 MV/cm) [3] and electron mobility (250-300 $cm^2$/Vs) [4] provide β-$Ga_2O_3$ a higher figure of merit than SiC and GaN. The availability of large area, high-quality bulk substrates from melt-growth methods [5,6,7,8] also provide a significant advantage for low-cost high-power devices. High voltage β-$Ga_2O_3$ transistors and Schottky barrier diode with various designs have been reported in recent years. [9,10,11,12,13,14,15,16,17] However, due to high acceptor activation energy and poor hole transport property, p-type β-$Ga_2O_3$ is not feasible for device design. The maximum breakdown field in vertical β-$Ga_2O_3$ devices is therefore limited by the Schottky barrier height to below 3.5 MV/cm. The breakdown of β-$Ga_2O_3$ MOS structure with dielectrics ($SiO_2$, $Al_2O_3$, $HfO_2$) is also limited by the metal/dielectric barrier height, and is seen to occur far below the material breakdown limit of β-$Ga_2O_3$.

We discuss the design principle of dielectric heterojunction diode. The schematics and simulated band diagram of the dielectric heterojunction diode are shown in Fig. 1 (a) (b) (c). It consists of a 20 nm high-k $BaTiO_3$ layer (eps ~300) on $Ga_2O_3$ lightly doped n-type drift layer. A Schottky metal contact is formed on the $BaTiO_3$ layer. $BaTiO_3$ is a perovskite material with a bandgap of 3.4 eV.

It has electron affinity of 3.8 eV [18], which is close to the value of β-Ga$_2$O$_3$. Thus, a small conduction band offset between BaTiO$_3$ and β-Ga$_2$O$_3$ can be expected. The Schottky barrier diode with the same β-Ga$_2$O$_3$ drift layer was also simulated as a comparison. Under reverse bias, the conduction band of the Schottky barrier diode (Fig.1 (d) (e) (f)) inclines sharply due to the electric field, resulting in a thin triangle electron barrier. As a result, the Schottky barrier diode shows breakdown at relatively low field (~3 MV/cm) due to Fowler-Nordhiem tunneling. When the BaTiO$_3$/ Ga$_2$O$_3$ dielectric heterojunction diode is reverse-biased, the electric field in the β-Ga$_2$O$_3$ drift layer is similar to the Schottky diode in the previous case. However, due to the large dielectric constant discontinuity between BaTiO$_3$ and β-Ga$_2$O$_3$, the electric field in BaTiO$_3$ is lower by a factor $\frac{\epsilon_{BTO}}{\epsilon_{GOX}}$ (~26 in this case) of the value in β-Ga$_2$O$_3$. As a result, the conduction band profile in BaTiO$_3$ remains flat, blocking the electrons tunneling from metal to β-Ga$_2$O under higher reverse bias. Therefore, the BaTiO3/ β-Ga$_2$O$_3$ dielectric heterojunction maintains a barrier at much higher voltages than the metal/semiconductor junction.

Under forward bias, electrons must flow from the semiconductor, through the high dielectric constant layer, into the metal. For small values of conduction band offset between BaTiO3/ β-Ga$_2$O$_3$ (as predicted from the electron affinity difference), dielectric heterojunction is expected to support efficient electron transport.

**Device fabrication**

To characterize the breakdown field of the dielectric heterojunction without additional field termination processes, we fabricated Schottky diodes and BaTiO$_3$/ Ga$_2$O$_3$ didoes on relative thin drift regions consisting of 150 nm unintentional doped (UID) β-Ga$_2$O$_3$ on n type substrate. The UID β-Ga$_2$O$_3$ drift layer was grown by oxygen plasma-assisted molecular beam epitaxy (MBE), using growth conditions reported earlier [19]. The electron concentration in UID β-Ga$_2$O$_3$ was estimated from capacitance-voltage profiling to be lower than $10^{16}$ cm$^{-3}$. 20 nm BaTiO$_3$ layer was deposited on β-Ga$_2$O$_3$ by physical vapor deposition in Ar/ O$_2$ ambient with a flow rate of 20: 2 sccm. Plasma power of 200 W and chamber pressure of 20 mTorr were maintained during deposition. A deposition rate of 0.5 nm/min was confirmed by ellipsometry. The sample surface before and after BaTiO$_3$ deposition were characterized by atomic force microscopy (AFM). (Fig. 2), and no significant difference in surface morphology was observed. A dielectric constant of 260±20 was measured from CV measurements. (The details were discussed in supplementary material.) A metal stack of Pt/Au (30 nm/100 nm) was deposited by electron beam evaporator to form square Schottky contact pads (50 um x 50 um) on both β-Ga$_2$O$_3$ sample and BaTiO$_3$/ β-Ga$_2$O$_3$ sample. 30 nm Ti followed by 100nm Au was deposited on the backside of n-type substrate to form Ohmic contact.

**Breakdown characteristics**

The reverse biased characteristics were measured on both the Schottky barrier diode and the BaTiO$_3$/ Ga$_2$O$_3$ dielectric heterojunction. (Fig. 3(a)) Using current of 1 A/cm$^2$ as breakdown condition, Schottky barrier diode had a breakdown voltage of 45 V, which corresponds to a breakdown field of 3 MV/cm, which is comparable with previous reports [14,15]. The BaTiO$_3$/Ga$_2$O$_3$ heterojunction diode showed significantly higher breakdown voltage of 85 V, corresponding to a peak electric field of 5.7 MV/cm (calculated assuming background doping = $10^{16}$ cm$^{-3}$). This is the highest breakdown electric field reported for a vertical β-Ga$_2$O$_3$ device. The 2-D electric field profile at breakdown was simulated using Sivalco ATLAS [20]. (Fig. 3(b) (c)) It shows when the electric field at the center of the device is 5.7 MV/cm, the edge of the device is expected to have an electric field of 7 MV/cm. The actual field in the devices fabricated may be higher than the simulated field, due to the square shape of the electrodes, which could cause field concentration at the edges.

We now discuss the breakdown mechanism of the $BaTiO_3/Ga_2O_3$ dielectric heterojunction. The Poole-Frenkel plot of the dielectric heterojunction (Fig. 4(a)) under reverse bias (40V - 70V) shows a linear behavior, indicating that the breakdown mechanism may governed by trap-assisted tunneling (Poole-Frenkel emission) [21]. The corresponding traps could located in either $BaTiO_3$ region or $Ga_2O_3$ region. The temperature dependent currents under reversed bias were measured and the Arrhenius plot regarding the electric field in $BaTiO_3$ and $Ga_2O_3$ were plotted separately in Fig. 4 (b) and (c). The trap energy level of 0.25 eV in $BaTiO_3$ and 0.6 eV in $Ga_2O_3$ were extracted from the slopes. The electric field used in this calculation was estimated in the center region of the device. However, the edges of the device have much higher electric field under reverse bias and they are likely to be the regions where breakdown happened. Therefore, the extract trap energy level may not be accurate. Further experiment and study is needed to confirm location and energy of the corresponding traps. Current results show the breakdown voltage of the dielectric heterojunction is not fundamentally limited by band structure, suggesting that improvement on the quality of $BaTiO_3$ and $Ga_2O_3$ may further increase the breakdown field of the dielectric heterojunction.

**Turn on characteristics**

Under forward bias, the Schottky junction and dielectric heterojunction showed on-resistance of 1.4 $m\Omega.cm^2$ and 1.5 $m\Omega.cm^2$, respectively (Fig. 5(a)). The 20 nm $BaTiO_3$ layer added 0.1 $m\Omega.cm^2$ extra on-resistance once the device is turned on. The turn-on voltage was 1.5 V for Schottky diode and 3.5 V for dielectric heterojunction diode respectively. The additional voltage drop is attributed to conduction band offset between BaTiO3 and β-Ga$_2$O. We found that the current-voltage characteristics can be explained using Fowler Nordheim [22] tunneling current analysis. The linear Fowler-Nordheim plot for the dielectric heterojunction diodes in forward bias condition suggests that the forward currents are limited by electron field emission across the BaTiO3 barrier (Fig. 5(b)). The effective barrier height was extracted to be 0.08 eV, which corresponds to the band offset between BaTiO$_3$ and Ga$_2$O$_3$. (Fig. 5(c))

**Conclusion**

In conclusion, we propose and demonstrate the dielectric heterojunctions that use extreme permittivity materials to achieve high breakdown field in a unipolar device. We demonstrate the BaTiO$_3$/Ga$_2$O$_3$ dielectric heterojunction diode with the record-high electric field of 5.7 MV/cm measured in Ga$_2$O$_3$. The dielectric heterojunction shows turn-on voltage of 3.5 V and on-resistance of 1.5 mΩ.cm$^2$. This structure demonstrated here not only brings up a new design for high performance Ga$_2$O$_3$ diode but also can be used as gate dielectric barrier in both lateral and vertical transistors. This dielectric heterojunction concept can also be expanded to other semiconductor materials in which bipolar device is technically challenging, such as SiC, GaN, ZnO and diamond. It may also be useful to Si and III-V materials to achieve faster unipolar device with higher breakdown voltage than typical Schottky barrier diode.

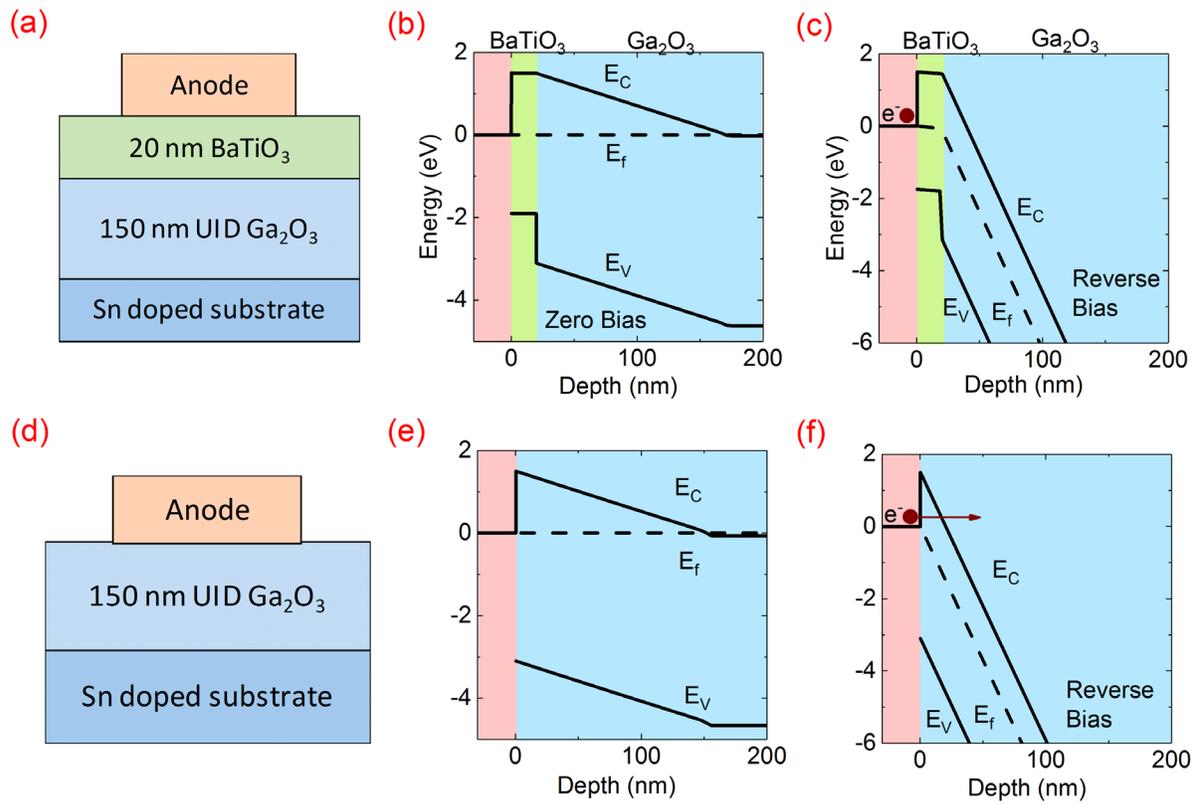

Fig. 1, (a) Schematics, (b) zero-biased band diagram and (c) reverse bias band diagram of BaTiO$_3$/β-Ga$_2$O$_3$ dielectric heterojunction. (d) Schematics, (e) zero-biased band diagram and (f) reverse bias band diagram of β-Ga$_2$O$_3$ Schottky barrier diode

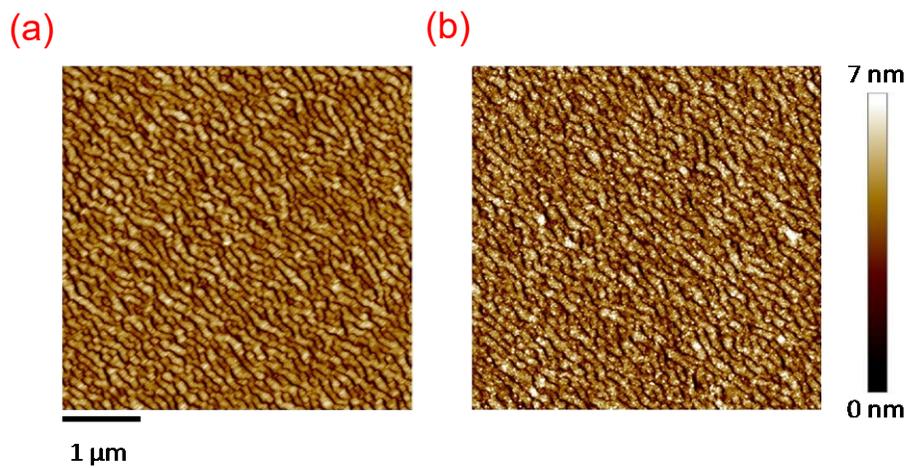

Fig. 2, The AFM image of (a) as-grown β-Ga$_2$O$_3$ surface and (b) 20 nm BaTiO$_3$ surface

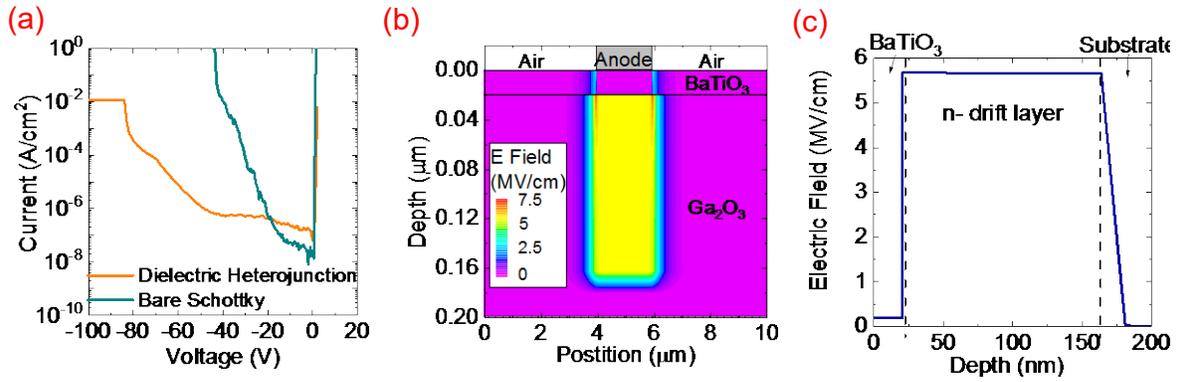

Fig. 3, (a) I-V characteristics under reverse biased (b) simulated 2-D electric field profile at breakdown bias (c) Simulated electric field profile from metal/ BaTiO$_3$ interface to n+ substrate

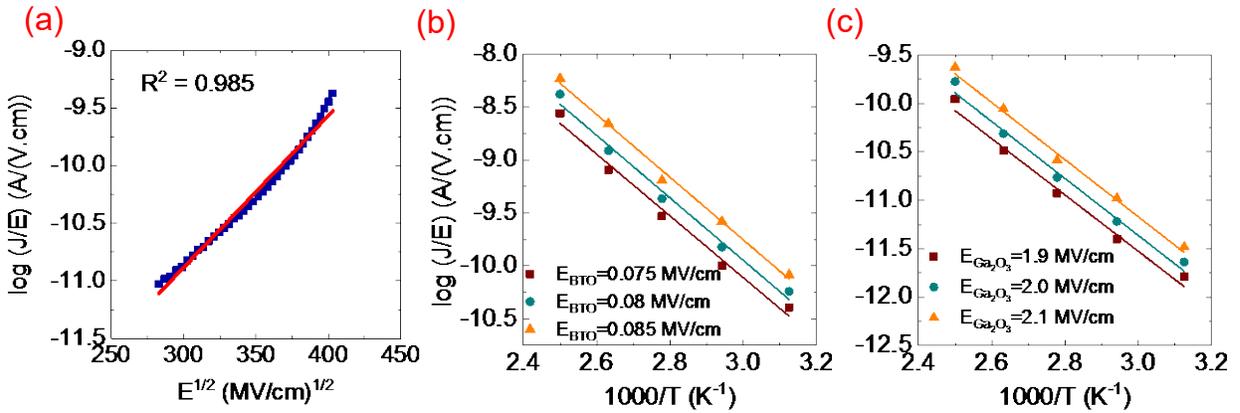

Fig. 4, (a) Poole-Frenkel plot of the dielectric heterojunction under reverse bais (40V-70V) and Arrhenius plot measured from 300 K -400 K regarding the electric field in (b) BaTiO$_3$ and (c) β-Ga$_2$O$_3$

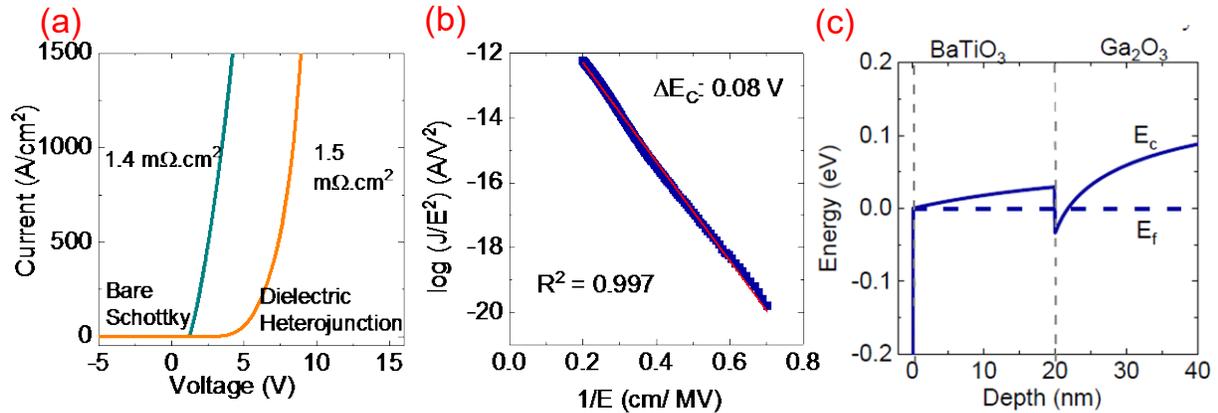

Fig. 5, (a) On- state characteristics of dielectric heterojunction diode and Schottky diode (b) Fowler-Nordhiem tunneling plot of the dielectric heterojunction diode under forward bias condition (c) Simulated conduction band diagram of dielectric heterojunction under forward bias condition (+1.5 V)

# Metal /BaTiO/ β-Ga2O3 Dielectric Heterojunction Diode with 5.7 MV/cm Breakdown Field

**Supplementary Information**

Zhanbo Xia[1], Hareesh Chandrasekar[1], Wyatt Moore[1], Caiyu Wang[1], Aidan Lee[3], Joe McGlone[1], Nidhin Kurian Kalarickal[1], Aaron Arehart[1], Steven Ringel[1,2], Fengyuan Yang[3], Siddharth Rajan[1,2]

[1] *Electrical & Computer Engineering, The Ohio State University, Columbus, OH 43210, USA*

[2] *Materials Science and Engineering, The Ohio State University, Columbus, OH 43210, USA*

3 *Department of Physics, The Ohio State University, Columbus, OH 43210, USA*

To determine the permittivity of the BaTiO$_3$ layer deposited by physical vapor deposition, two samples with delta-doped structure on Sn-doped substrate was used for capacitance-voltage measurement. The test structure consists a Si delta doped layer with 1.5 x10$^{13}$ cm$^{-2}$ doping concentration on a 60 nm UID Ga$_2$O$_3$ buffer. A 20 nm UID Ga$_2$O$_3$ cap layer was grown on the Si delta doped layer. The use of Si delta doped layer near the surface is to form a capacitance that is comparable to 20 nm BaTiO$_3$ so the capacitance of BaTiO$_3$ can be easily differentiated. A 20nm BaTiO$_3$ layer was deposited on one of the samples by physical vapor deposition (sample a). (Fig.1 (a))The other remains as-grown surface as a control sample (Sample b). (Fig.1 (b))The simulated band diagram and electron density profile of the control sample (sample b) are shown in Fig. 2. Schottky contact and Ohmic contact were formed on the top and bottom of the samples respectively. C-V characteristic was measured on both sample shown in Fig. 3 (a). The electron density profile showing a 2-D electron gas was calculated from the measured C-V curve shown in Fig. 3 (b). The 2DEG density peak indicate the position of Si delta doping site. The capacitances (C$_a$ for sample A and C$_b$ for sample B) associated with the peak point for the two samples were

extracted respectively. The capacitance associated with the 20 nm BaTiO$_3$ layer can be calculated by:

$$C_{BTO} = \frac{C_a C_b}{C_b - C_a}$$

The dielectric constant of 260 ± 20 was calculated based the capacitance of the 20 nm BaTiO$_3$ layer.

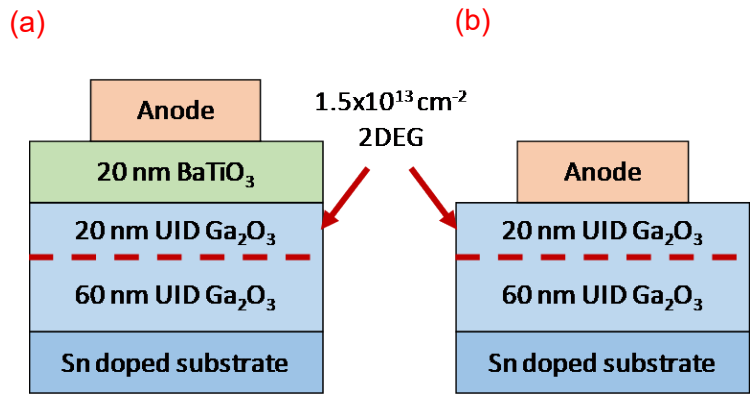

Fig. 1, Schematics of delta-doped sample (a) with BaTiO$_3$ layer and (b) without BaTiO$_3$ layer

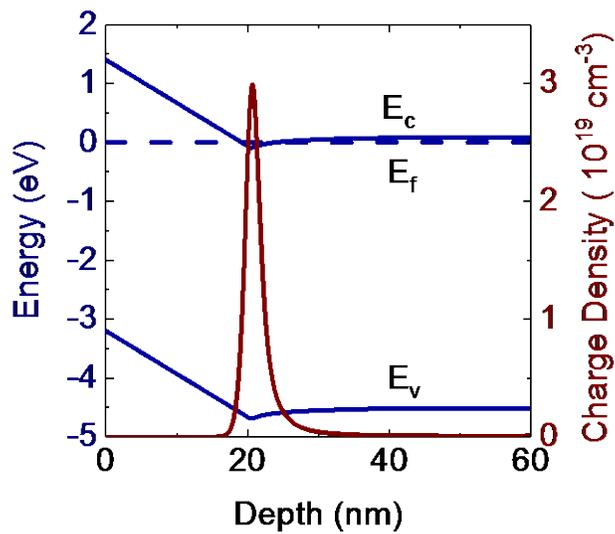

Fig. 2 Simulated band diagram and electron density profile of the delta-doped sample

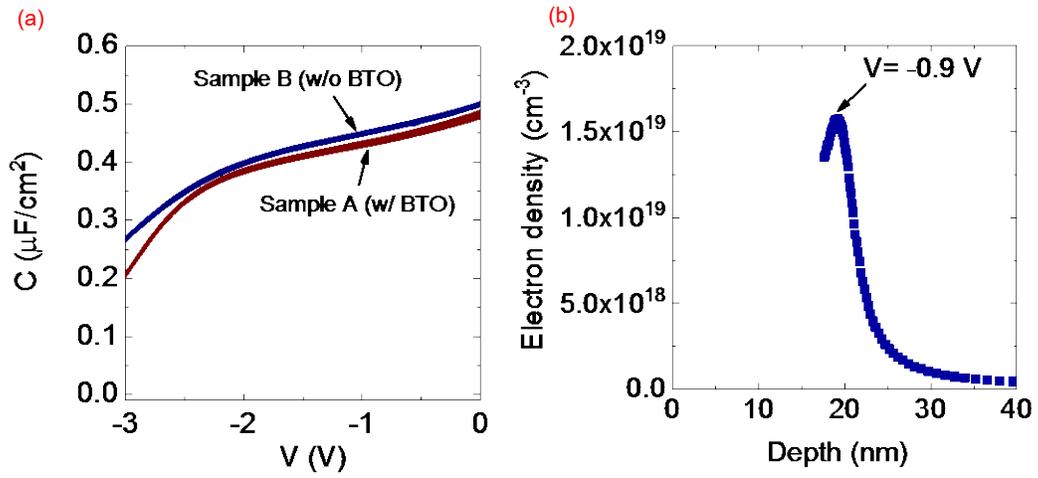

Fig. 3 (a) Measured C-V profile of sample a and sample b (b) extracted electron density profile with peak density measured